\documentclass[letterpaper,twocolumn,english,aps,final,superscriptaddress,nobibnotes]{revtex4-1}
\usepackage[T1]{fontenc}
\usepackage[latin9]{inputenc}
\usepackage{color}
\usepackage{babel}
\usepackage{amsmath}
\usepackage{amssymb}
\usepackage{graphicx}
\usepackage{esint}
\usepackage[unicode=true,pdfusetitle,
 bookmarks=true,bookmarksnumbered=false,bookmarksopen=false,
 breaklinks=false,pdfborder={0 0 1},backref=false,colorlinks=false]
 {hyperref}
\hypersetup{
 colorlinks=true, linkcolor=blue, urlcolor=blue, citecolor=blue, pdfstartview={FitH}, hyperfootnotes=false, unicode=true}
\usepackage{breakurl}

\makeatletter

\usepackage{babel}
\usepackage{dcolumn}
\usepackage{bm}

\usepackage{times}

\setkeys{Gin}{width=\ifdim 0.75\Gin@nat@width>\linewidth
  \linewidth
\else
  0.75\Gin@nat@width
\fi}

\makeatother

\begin{document}

\title{Giant electrocaloric response in the prototypical Pb(Mg,Nb)O$_{3}$
relaxor ferroelectric \\
 from atomistic simulations}

\author{Zhijun Jiang}

\affiliation{School of Microelectronics and State Key Laboratory for Mechanical
Behaviour of Materials, Xi'an Jiaotong University, Xi'an 710049, China}

\affiliation{Physics Department and Institute for Nanoscience and Engineering,
University of Arkansas, Fayetteville, Arkansas 72701, USA }

\author{Y. Nahas}

\affiliation{Physics Department and Institute for Nanoscience and Engineering,
University of Arkansas, Fayetteville, Arkansas 72701, USA }

\author{S. Prokhorenko}

\affiliation{Physics Department and Institute for Nanoscience and Engineering,
University of Arkansas, Fayetteville, Arkansas 72701, USA }

\affiliation{Theoretical Materials Physics, Q-MAT CESAM, University of Liège,
B-4000 Sart Tilman, Belgium}

\author{S. Prosandeev}

\affiliation{Physics Department and Institute for Nanoscience and Engineering,
University of Arkansas, Fayetteville, Arkansas 72701, USA }

\affiliation{Research Institute of Physics and Physics Department, Southern Federal
University, Rostov-on-Don 344090, Russia}

\author{D. Wang}

\affiliation{School of Microelectronics and State Key Laboratory for Mechanical
Behaviour of Materials, Xi'an Jiaotong University, Xi'an 710049, China}

\author{Jorge Íñiguez}

\affiliation{Materials Research and Technology Department, Luxembourg Institute
of Science and Technology, 5 avenue des Hauts-Fourneaux, L-4362 Esch/Alzette,
Luxembourg}

\author{L. Bellaiche}

\affiliation{Physics Department and Institute for Nanoscience and Engineering,
University of Arkansas, Fayetteville, Arkansas 72701, USA }
\begin{abstract}
An atomistic effective Hamiltonian is used to investigate electrocaloric
(EC) effects of Pb(Mg$_{1/3}$Nb$_{2/3}$)O$_{3}$ (PMN) relaxor ferroelectrics
in its ergodic regime, and subject to electric fields applied along
the pseudocubic {[}111{]} direction. Such Hamiltonian qualitatively
reproduces (i) the electric field-\textit{versus}-temperature phase
diagram, including the existence of a critical point where first-order
and second-order transitions meet each other; and (ii) a giant EC
response near such critical point. It also reveals that such giant
response around this critical point is microscopically induced by
field-induced percolation of polar nanoregions. Moreover, it is also
found that, for any temperature above the critical point, the EC coefficient-\textit{versus}-electric
field curve adopts a maximum (and thus larger electrocaloric response
too), that can be well described by the general Landau-like model
proposed in {[}Jiang \textit{et al}, Phys. Rev. B \textbf{96}, 014114
(2017){]} and that is further correlated with specific microscopic
features related to dipoles lying along different rhombohedral directions.
Furthermore, for temperatures being at least 40 K higher than the
critical temperature, the (electric field, temperature) line associated
with this maximal EC coefficient is below both the Widom line and
the line representing percolation of polar nanoregions. 
\end{abstract}
\maketitle

\section{Introduction}

The electrocaloric (EC) effect characterizes the change in dipolar
entropy or temperature under the application and/or removal of an
electric field \cite{Lines1997,Jona1993,Scott2011,Zhang2006,Zhang_APL_2006,Kutnjak2015}\textcolor{black}{.}
It has the potential to lead to the design of efficient solid-state
cooling devices for a broad range of applications \cite{Bai2010,Moya2014,Kutnjak2015,Zhang2014}.
As such, EC effects have been intensively studied in recent years
(see, e.g., Refs. \cite{Liu2013,Liu2014,Sanlialp2015,Lines1997,Kutnjak2015,Uchino2000,Prosandeev2008,Lisenkov2009,Ponomareva2012,Rose2012,Marathe2016,Geng2015,Defay2013,Defay2016,Defay_2016,Guzman-Verri2016,Marathe2014,Pirc2011-1,Marathe2017}
and references therein). In particular, a promising large electrocaloric
response has been measured in prototypical lead-based relaxor ferroelectrics,
such as Pb(Mg,Nb)O$_{3}$ (PMN), (Pb,La)(Zr,Ti)O$_{3}$ and Pb(Mg,Nb)O$_{3}$\textendash PbTiO$_{3}$
\cite{Pirc2011-1}, in the vicinity of the critical point where first-order
and second-order transitions meet in the electric field-\textit{versus}-temperature
phase diagram. Relaxor ferroelectrics differentiate themselves from
typical ferroelectrics, by, e.g., adopting a frequency-dependent dielectric
response-\textit{versus}-temperature function, as well as several
characteristic temperatures \cite{Burns1983,Dkhil2009,Svitelskiy2005,Vogel1921,Fulcher1925,Jeong2005,Nahas2016}
even if they remain macroscopically paraelectric down to 0 K. It is
important to realize that two types of relaxor ferroelectrics should
be distinguished because they can exhibit different properties: Pb-based
ones, such as PMN, \textit{versus} lead-free ones, such as Ba(Zr$_{0.5}$Ti$_{0.5}$)O$_{3}$
(BZT). For instance, unlike PMN, there is no aforementioned critical
point present in BZT. Another evidence of their possible difference
is that the relaxor nature of BZT was predicted to originate from
small Ti-rich polar nanoregions (PNRs) as a result of the difference
in polarizability between Ti and Zr sites \cite{Akbarzadeh2012},
while the lead-based PMN system was numerically found to be a relaxor
because of a complex interplay between random electric fields, ferroelectric
and antiferroelectric interactions \textendash{} with such interplay
yielding much larger PNRs touching each other at low temperatures
\cite{Prosandeev2015}.

Due to its complexity, unlike typical ferroelectrics \cite{Prosandeev2008,Lisenkov2009,Ponomareva2012,Rose2012,Marathe2016}
and lead-free relaxor ferroelectrics \cite{Jiang2017}, we are not
aware of any \textit{atomistic} simulation devoted to the study of
EC effects in lead-based relaxors. Consequently, several questions
remain unanswered in systems such as PMN. For instance, are atomistic
modeling able to reproduce the existence of a critical point in such
complex compound and reveal atomistic features (if any) associated
with the enhancement of EC response near the critical point? In particular,
could such features be related to electric-field-induced percolation
of the polar nanoregions? It is also legitimate to wonder if, for
temperatures higher than the critical point, some electric fields
can also yield an enhancement of the EC response (i.e., a large electrocaloric
response too), and if such enhancement can be traced back to specific
atomistic features? In addition, while the simple Landau-type phenomenological
model developed in Ref. \cite{Jiang2017} can reproduce the temperature-
and field-driven behavior of the EC response of typical ferroelectrics
and lead-free relaxor ferroelectrics, it is important to determine
if such model is also valid in the more complex PMN compound, which
will make such model even more general and of broader use (note that
such a model predicts that the EC coefficient is directly related
to the product of the temperature and the derivative of the \textit{square}
of the polarization with respect to electric field).

The goal of this article is to provide an answer to all the aforementioned
questions in the PMN relaxor ferroelectric subject to dc electric
fields applied along the pseudocubic {[}111{]} direction. For that,
we will adopt the following organization. Section II provides details
about the atomistic method employed here, as well as our practical
way to compute the EC response. Section III.A demonstrates that such
atomistic method is indeed able to qualitatively reproduce the peculiar
(electric field, temperature) phase diagram of PMN, including its
critical point. Section III.B shows that (1) there is indeed an electric
field leading to a maximal EC coefficient for any fixed temperature
above the critical point, with such maximal EC coefficient being strongly
enhanced when decreasing the temperature such as to approach the critical
point from above; and (2) that the Landau-type model of Ref. \cite{Jiang2017}
is still rather accurate for PMN. Section III.C is dedicated to local
atomistic features inherently linked to EC responses. In particular,
the giant EC coefficient numerically found in the vicinity of the
critical point is revealed to be correlated with field-induced percolation
of polar nanoregions, while the optimization of the EC response for
higher temperature is linked to other, subtle and original microscopic
characteristics. Finally, Section IV summarizes this work.

\section{Methods}

Here, we use the first-principles-based effective Hamiltonian ($H_{\textrm{eff}}$)
approach developed in Ref. {[}\onlinecite{Prosandeev2015}{]}. Its
total internal energy contains two main terms, $E_{int}(\{\mathrm{\mathbf{u}}_{i}\},\thinspace\{\mathbf{v}_{i}\},\thinspace\eta_{H},\thinspace\{\sigma_{j}\})=E_{\mathrm{ave}}(\{\mathrm{\mathbf{u}}_{i}\},\thinspace\{\mathbf{v}_{i}\},\thinspace\eta_{H})+E_{\mathrm{loc}}(\{\mathrm{\mathbf{u}}_{i}\},\thinspace\{\mathbf{v}_{i}\},\thinspace\{\sigma_{j}\})$,
where $\{\mathrm{\mathbf{u}}_{i}\}$ is the Pb-centered local soft
mode in unit cell $i$ (which is proportional to the electric dipole
moment of that cell), $\{\mathbf{v}_{i}\}$ are variables related
to the inhomogeneous strain and are centered on the B sites (Ng or
Mg ions), $\eta_{H}$ is the homogeneous strain tensor, and $\{\sigma_{j}\}$
characterizes the atomic distribution of Mg and Nb ions. $E_{\mathrm{ave}}$
describes the energies of a simple virtual perovskite system and has
five terms: (i) the local-mode self-energy; (ii) the long-range dipole-dipole
interaction; (iii) the short-range interactions between local modes;
(iv) the elastic energy; and (v) the energy representing the interaction
between local modes and strains \cite{Zhong1995}. $E_{loc}$ mimics
how the distribution of Mg and Nb cations alters energetics \cite{Prosandeev2015}.
We also add to $E_{int}$ an energy that is proportional to minus
the dot product between polarization and electric field, in order
to simulate the effect of such field on properties.

We employ this $H_{\textrm{eff}}$ within Monte Carlo (MC) simulations
on $18\times18\times18$ supercells (29,160 atoms) with periodic boundary
conditions. Mg and Nb ions are randomly distributed inside these supercells.
20,000 MC sweeps are used for equilibration and an additional 80,000
MC sweeps are employed to compute statistical averages at desired
temperature, $T$, and electric field, ${\cal E}$, in order to get
converged results. Typically and unless specified in figures' captions,
we use here one disordered chemical configuration, in order to capture
the first-order nature of some electric-field-driven transitions (since
different random arrangements can have slightly different critical
fields for these first-order transitions, and therefore averaging
over different configurations will, e.g., smear out the first-order-induced
jump of the polarization when increasing the electric field).

The EC coefficient, $\alpha$, is defined to be the derivative of
the temperature with respect to electric field at constant entropy,
and is computed from MC runs via the following cumulant formula \cite{Omran2016,Jiang2017}:
\begin{equation}
\alpha=-\thinspace Z^{*}a_{lat}NT\thinspace\{\frac{\left\langle \mathbf{\left|u\right|}{E_{int}}\right\rangle -\left\langle \mathbf{\left|u\right|}\right\rangle \left\langle {E_{int}}\right\rangle }{\left\langle {E_{int}}^{2}\right\rangle -\left\langle {E_{int}}\right\rangle ^{2}}\},\label{eq:ece-alpha-cumulant}
\end{equation}
where $Z^{*}$ is the Born effective charge, $a_{lat}$ is the five-atom
lattice constant, $N$ is the number of sites in the supercell, $T$
is the selected temperature, $\mathbf{u}$ is the supercell average
of the local mode, ${E_{int}}$ is the total internal energy of the
$H_{\textrm{eff}}$, and $\left\langle \ \right\rangle $ denotes
the average over the MC sweeps at every selected temperature. The
computation of $\alpha$ via Eq. (\ref{eq:ece-alpha-cumulant}) is
done for a chosen combination of temperature and magnitude of a dc
electric field applied along the pseudocubic {[}111{]} direction.

\section{Results }

\subsection{${\cal E}$-$T$ phase diagram}

\begin{figure}
\includegraphics[width=7cm]{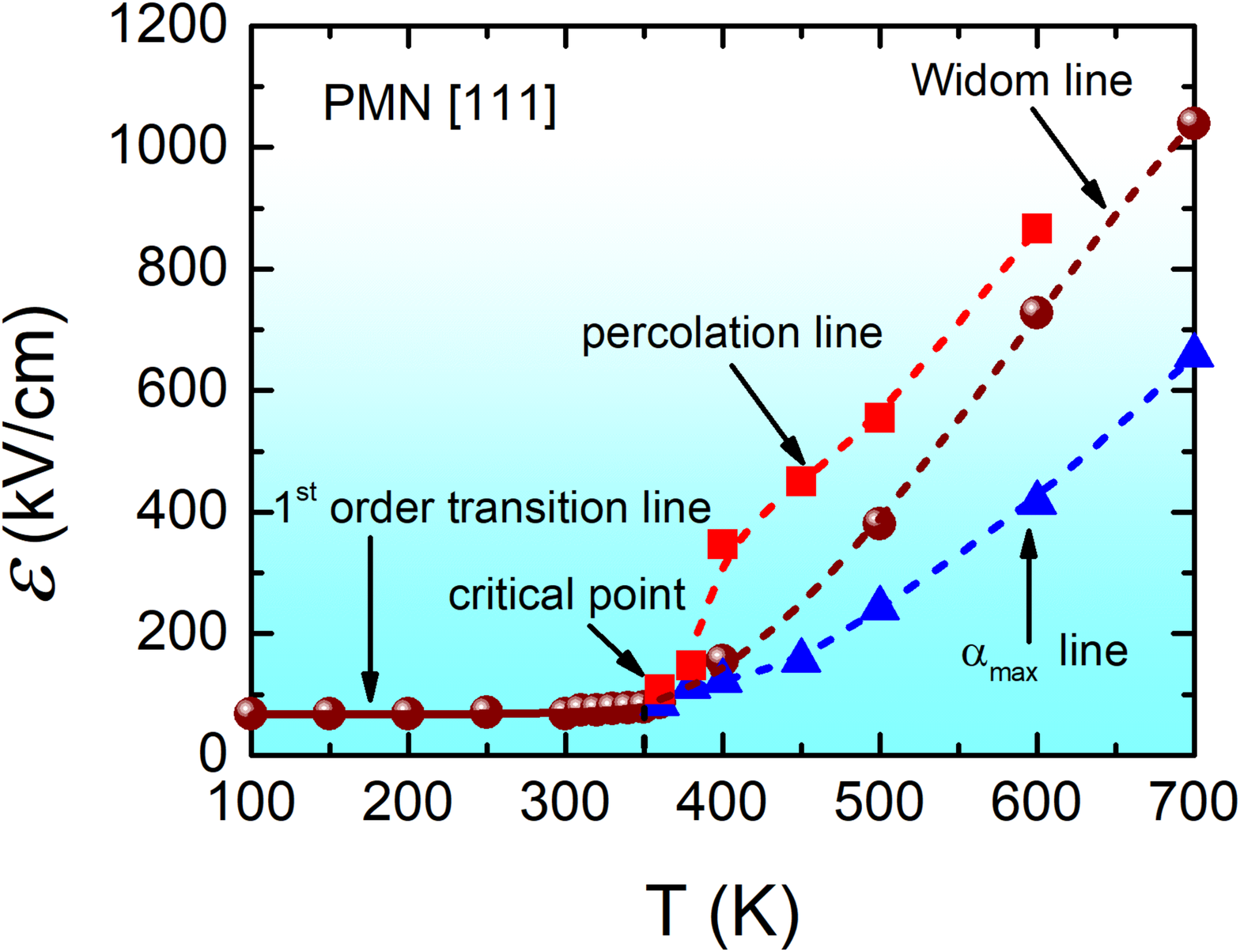}

\caption{${\cal E}$-$T$ phase diagram of PMN for dc electric fields applied
along the pseudocubic {[}111{]} direction, as predicted by our $H_{\textrm{eff}}$
when varying the magnitude of the electric field for each fixed, considered
temperature. The solid line represents first-order transitions between
non-ergodic and ferroelectric states, while the brown dashed line
displays the Widom line. These two lines meet at the critical (${\cal E_{CP}}$,$T_{CP}$)
point. Two additional dashed lines are indicated in this figure: the
blued one along which the EC $\alpha$ coefficient is maximum for
any considered temperature above $T_{CP}$, and the red one that displays
the location of percolation for $T$ $\ge$ $T_{CP}$. \label{fig:Phase diagram}}
\end{figure}

Let us start by determining the ${\cal E}$-$T$ phase diagram of
PMN, as predicted from the use of our $H_{\textrm{eff}}$ for a given
disordered configuration. Figure \ref{fig:Phase diagram} shows such
phase diagram, when varying the magnitude of the electric field along
the {[}111{]} direction while keeping the temperature constant (for
different choices of this temperature ranging between 100 and 700
K). Two different particular lines can be seen there: (1) a solid
line corresponding to a first-order transition from a non-ergodic
relaxor state to a ferroelectric state, as consistent with measurements
\cite{Zhao2007,Kutnjak2008,Kutnjak2007} and as numerically found
via the occurrence of a sudden jump in the polarization-versus-${\cal E}$
curve at fixed temperature (see Supplemental Material \cite{SM});
and (2) a brown dashed line corresponding to the so-called Widom line
\cite{Xu2005,Simeoni2010,Kutnjak2007} and that is presently identified
via the occurrence of peaks in the dielectric response (see Supplemental
Material \cite{SM}, while the polarization-versus-${\cal E}$ function
is continuous). Interestingly, these two lines meet at a critical
point to be denoted as (${\cal E_{CP}}$, $T_{CP}$) and which is
equal to (86.6 kV/cm, $\simeq$ 360 K). Our predicted phase diagram
of Fig. \ref{fig:Phase diagram} therefore qualitatively agrees with
those measured in Refs. \cite{Kutnjak2007,Kutnjak2008,Zhao2007},
that also exhibit a critical point, along with a first-order transition
line below $T_{CP}$ and a Widom line above $T_{CP}$. Quantitatively,
our simulated ${\cal E_{CP}}$ is about 22 times larger than the measured
one \cite{Kutnjak2007}, which is typical for atomistic simulations
\cite{Jiang2017,Xu2017}, while the resulting predicted $T_{CP}$
is about 130 K higher than the observed one of 230 K \cite{Kutnjak2008}.

\subsection{EC coefficients}

\begin{figure}
\includegraphics[width=7cm]{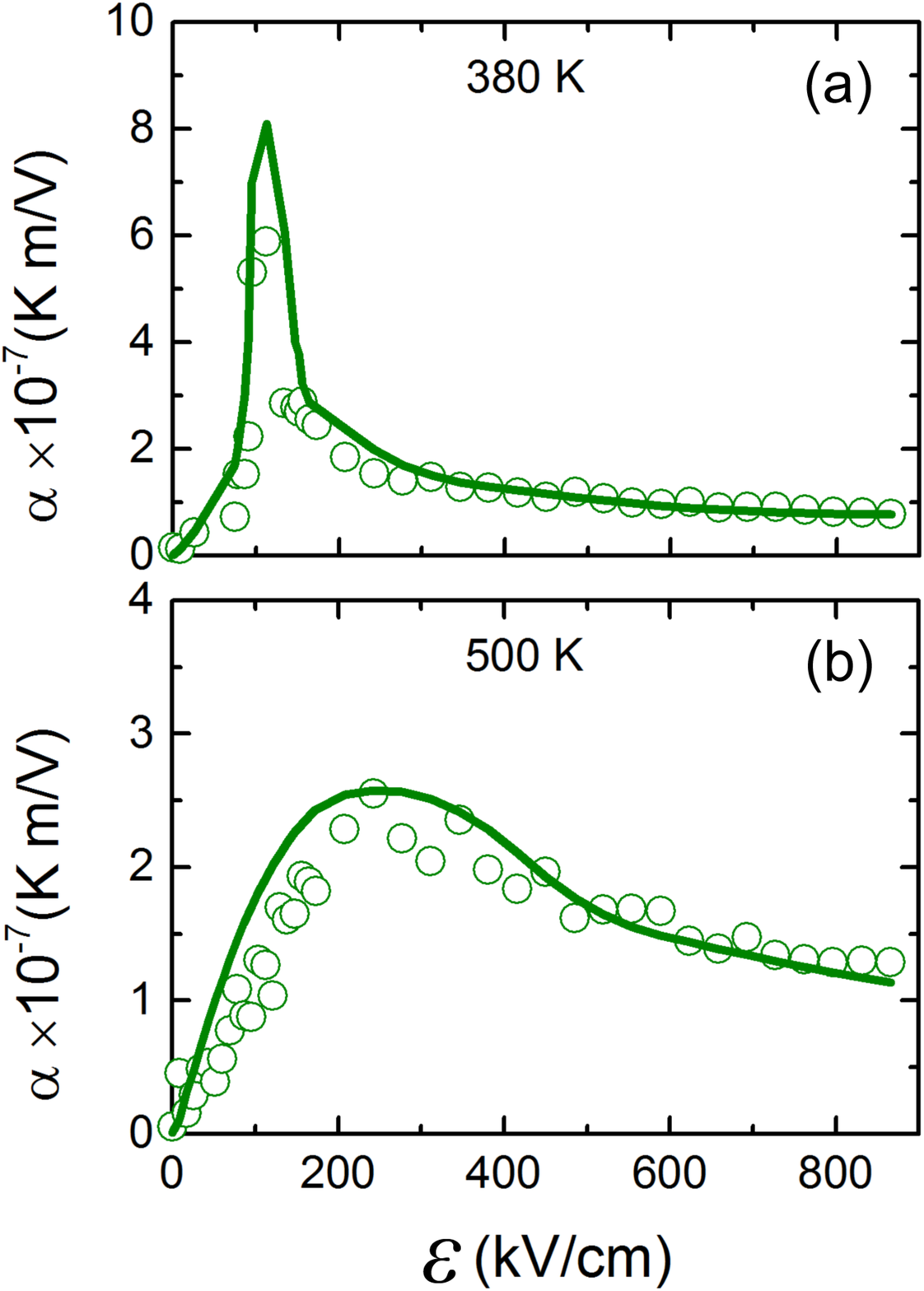}

\caption{Electrocaloric coefficient, $\alpha$, as a function of the applied
dc electric field ${\cal E}$, at (a) 380 K and (b) 500 K. The solid
green lines represent the fit of the MC results by the Landau-like
model of Ref. \cite{Jiang2017}, i.e., $\alpha=\beta T\left.\frac{\partial P^{2}}{\partial{\cal E}}\right|_{T}$,
where $\beta$ is a constant. \label{fig:alpha_vs_E}}
\end{figure}

Let us now concentrate on the EC coefficient. It is important to recall
that Eq. (\ref{eq:ece-alpha-cumulant}) automatically assumes ergodic
conditions. Since such conditions are ``only'' satisfied for temperatures
above $T_{CP}$ for any field in the phase diagram of Fig. \ref{fig:Phase diagram}
(recall that for $T<T_{CP}$ and ${\cal E}<{\cal E}_{CP}$, the system
is nonergodic), we decided to limit the present investigation of EC
effects in PMN for temperatures equal or higher than $\simeq360$
K. Figure \ref{fig:alpha_vs_E} shows the electrocaloric coefficient
as a function of electric field, ${\cal E}$, for two selected temperatures,
namely 380 and 500 K (that therefore both lie in the ergodic regime).
For any presently investigated temperature, $\alpha$ exhibits a non-monotonic
behavior with field that has also been previously seen in the lead-free
Ba(Zr,Ti)O$_{3}$ relaxor ferroelectric \cite{Jiang2017}. Such behavior
consists of vanishing values at low fields, followed by an increase
up to a maximum (to be denoted as $\alpha_{max}$) before decreasing
for larger fields.

\begin{figure}
\includegraphics[width=7cm]{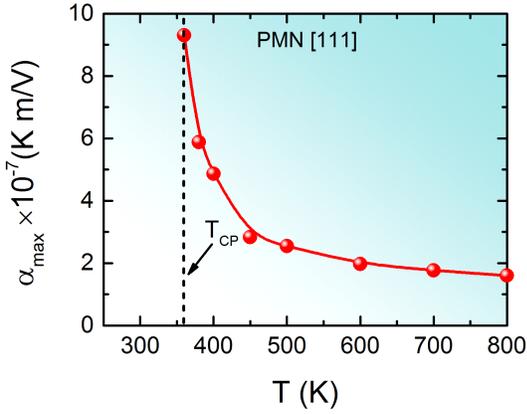}

\caption{Maximal value of the electrocaloric coefficient, $\alpha_{max}$,
as a function of temperature. \label{fig:alpha_max}}
\end{figure}

Moreover, Fig. \ref{fig:alpha_max} reports $\alpha_{max}$ as a function
of temperature. It is clear that, in the ergodic regime, $\alpha_{max}$
increases when the temperature decreases down to the critical point
$T_{CP}\backsimeq360$ K, which is in qualitative agreement with experimental
data of PMN \cite{Pirc2011-1} and which emphasizes the importance
of proximity to the critical point for the enhancement of the electrocaloric
effect. Interestingly, our predicted value of $\alpha_{max}$ at 380
K is of the order of $6.0\times10^{-7}$ K m/V, that is of the same
order than the experimental data of $3.0\times10^{-7}$ K m/V at the
measured $T_{CP}$ critical temperature of PMN \cite{Pirc2011-1}.
Note that $\alpha_{max}$ is still large at, e.g., 500 K, since it
is computed to be of the order of $2.0\times10^{-7}$ K m/V.

Furthermore, Fig. \ref{fig:Phase diagram} further displays the value
of the specific electric field at which $\alpha$ is maximum for any
investigated temperature above $T_{CP}$. It reveals that, for any
of these temperatures (at the sole exception of $T_{CP}$), this field
is lower than that of the Widom line. Such feature can be understood
by the fact that, as previously found for Ba(Zr,Ti)O$_{3}$ relaxor
ferroelectrics as well as for prototypical ferroelectrics \cite{Jiang2017}
and as shown in Figs. \ref{fig:alpha_vs_E}(a) and \ref{fig:alpha_vs_E}(b)
by means of solid green curves, the behavior of $\alpha$ versus electric
field for any considered temperature is found here to be very well
reproduced by a simple Landau-derived model (note that Ref. \cite{Jiang2017}
provides more details about this model, assumptions and the resulting
derived final formula) indicating that $\alpha$ should be equal to
$\beta T\left.\frac{\partial P^{2}}{\partial{\cal E}}\right|_{T}$,
where $\beta$ is a constant and $P$ is the polarization. Such fact
further demonstrates the generality of such simple model, and the
intrinsic relationship between the EC coefficient and the derivative
of the \textit{square} of the polarization with respect to electric
field at constant temperature. The electric field leading to the enhancement
of $\alpha_{max}$ at a fixed temperature is therefore not the one
of the Widom line because this latter is related to the vanishing
of the derivative of the dielectric constant with respect to temperature
at fixed electric field (which thus leads to the annihilation of the
second derivative of the polarization with respect to both electric
field and temperature) rather than the vanishing of the second derivative
of the square of the polarization with respect to electric field at
constant temperature (which is the case for $\alpha_{max}$).

Let us now check if the electric fields associated with $\alpha_{max}$
can be rather traced back to local features.

\subsection{Local features}

\begin{figure}
\includegraphics[width=7cm]{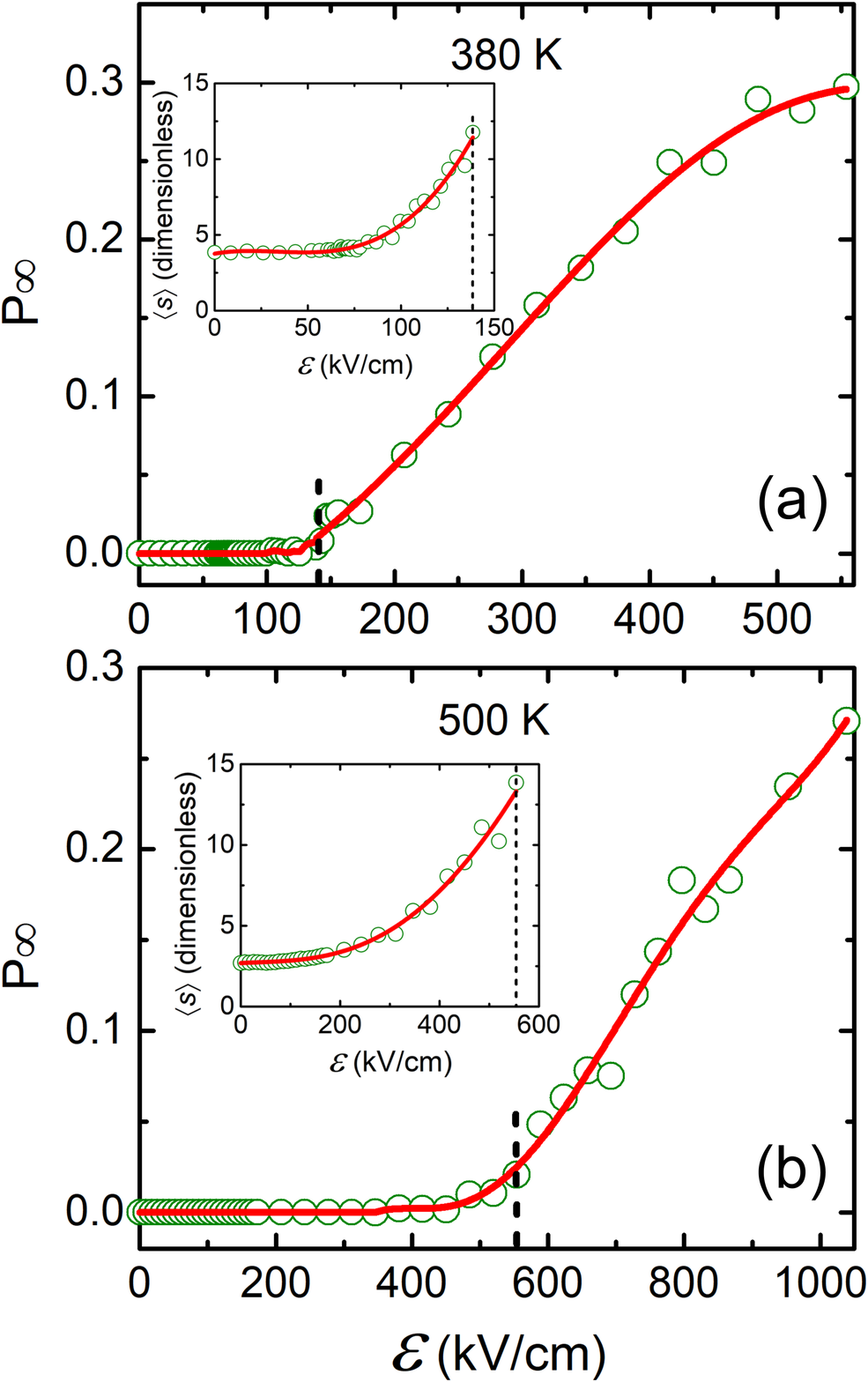}

\caption{Dependency of the strength of the percolating cluster on the magnitude
of the electric field applied along {[}111{]} in disordered PMN solid
solutions, at 380 K (Panel a) and 500 K (Panel b). The insets show
the average cluster size as a function of field. $P_{\infty}$ and
$\left\langle s\right\rangle $ are averaged here over 20 different
disordered PMN configurations, in order to obtain a better statistics.
The red lines are guides for the eye. \label{fig: percolation}}
\end{figure}

For that, we first decided to resort to percolation theory and computed
two specific quantities. The first one is the so-called strength of
the percolating cluster \cite{Stauffer1994,Prosandeev2013}, that
is calculated as $P_{\infty}=N_{\infty}/N_{\mathit{\textrm{Pb}}}$,
where $N_{\infty}$ is the number of the distinct Pb sites of the
supercell belonging to the (infinite) percolating cluster {[}note
that the infinite cluster is defined to be a cluster spreading from
one side of the supercell to the opposite side, and inside which the
dipoles are nearly parallel to each other (that is, when the cosine
of the angle between two nearest neighboring dipoles is larger than
0.85){]} and where $N_{\textrm{Pb}}$ is the number of Pb ions in
the whole supercell. The second quantity is the average cluster size
\cite{Stauffer1994,Prosandeev2013,Prosandeev2015}, which is computed
as $\left\langle s\right\rangle =\left\langle N^{2}\right\rangle /\left\langle N\right\rangle $,
where $N$ is the number of Pb sites belonging to a polar nanoregion,
and the brackets denote the average over all the PNRs existing inside
the supercell (note that the criterion presently used to numerically
find if two dipoles centered on first nearest-neighbors Pb ions belong
to the same PNR is that the angle between these two dipoles has a
cosine being between 0.85 and 1.0). Note that $\left\langle s\right\rangle $
is only computed here when the strength of the percolating cluster
is negligible, since $\left\langle s\right\rangle $ is only physical
when the percolating cluster has not formed yet.

Figures \ref{fig: percolation}(a) and \ref{fig: percolation}(b)
show the strength of the percolating cluster as a function of the
magnitude of the electric field at 380 and 500 K, respectively, with
their insets displaying the corresponding field dependency of the
average cluster size at these two temperatures. At 380 K, $P_{\infty}$
basically vanishes below ${\cal E\backsimeq}$ 140 kV/cm, and then
becomes finite and significantly increases when the field further
increases. Moreover, the inset of Fig. \ref{fig: percolation}(a)
reveals that \textbf{$\left\langle s\right\rangle $} is nearly constant,
around 4, for fields below 87 kV/cm, and then is rapidly enhanced
when ${\cal E}$ increases up to 140 kV/cm. Such behaviors imply that
the PNRs are first typically small for low fields and then rapidly
become bigger for larger fields, until they percolate at the specific
field of 140 kV/cm for the temperature of 380 K. Strikingly, such
percolating field of 140 kV/cm is very close to the value of the field
at which $\alpha$ adopts its maximal value at 380 K {[}see Fig. \ref{fig:alpha_vs_E}(a){]}.
In other words, our results reveal that, close to the critical point
(${\cal E_{CP}}$,$T_{CP}$), the EC coefficient is optimized when
percolation of dipoles occurs at the atomistic scale. To know if such
fact also holds for higher temperature, one can now pay attention
to the data of Fig. \ref{fig: percolation}(b) corresponding to 500
K. In that case, the percolating field is close to $\backsimeq$ 554
kV/cm, which is larger than the field of $\backsimeq$290 kV/cm at
which $\alpha$ is maximum at 500 K {[}see Fig. \ref{fig:alpha_vs_E}(b){]}
(note also that the average cluster size at low fields is now close
to 2.5 at 500 K {[}see the inset of Fig. \ref{fig: percolation}(b){]},
which is smaller than 4 at 380 K, and which explains why one needs
larger fields to induce percolation at larger temperature). In other
words, percolating fields are not necessarily the fields at which
the EC coefficient is optimal for any temperature above $T_{CP}$.
In fact, and as also demonstrated by Fig. \ref{fig:Phase diagram}
that further reports the fields at which percolation occurs for temperatures
above $T_{CP}$, it is only for temperatures lying between $\simeq$
360 and 400 K (that is near the predicted value \cite{Prosandeev2015}
of the so-called $T^{*}$ of PMN \cite{Viehland1990,Dkhil2001,Stock2010})
that the field yielding a maximum of $\alpha$ is close to the percolating
field.

\begin{figure}
\includegraphics[width=7cm]{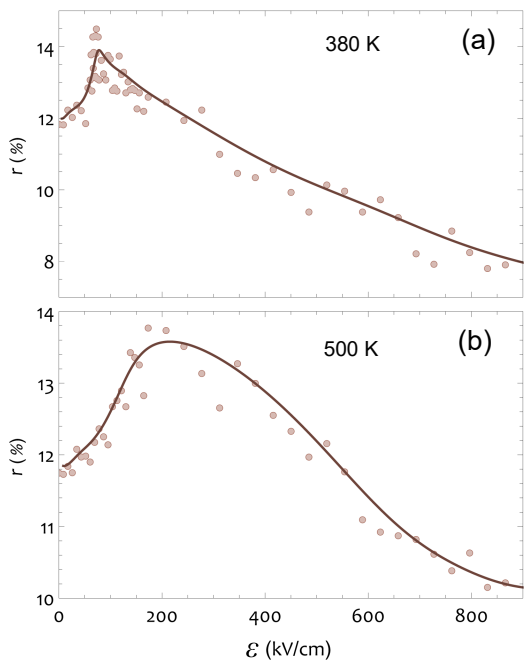}

\caption{Percentage of dipoles lying near the {[}1$\overline{1}$1{]}, {[}11$\overline{1}${]}
or {[}$\overline{1}$11{]} pseudocubic direction, as a function of
the magnitude of the dc electric field applying along the {[}111{]}
direction, at (a) 380 K and (b) 500 K. \label{fig: microscopic}}
\end{figure}

Let us thus now search for other local features that can better correlate
with the enhancement of the EC coefficient for both 380 and 500 K.
For that, we computed the percentage of dipoles in the supercell that
lie near (namely, within 25$^{\circ}$) the {[}1$\overline{1}$1{]},
{[}11$\overline{1}${]} or {[}$\overline{1}$11{]} pseudocubic directions,
as a function of the magnitude of the dc electric field (that, we
recall, is applied along {[}111{]}). In other words, we numerically
determined the percentage of dipoles lying near all the rhombohedral
directions that have a positive projection on the applied field, at
the sole exception of this applied {[}111{]} direction. Figures \ref{fig: microscopic}(a)
and \ref{fig: microscopic}(b) show such percentage at 380 and 500
K, respectively, and reveal that it exhibits a maximum at some specific
temperature-dependent field. Interestingly, such latter fields are
basically those associated with the maximal values of $\alpha$ at
380 and 500 K {[}see Figs. \ref{fig:alpha_vs_E}(a) and \ref{fig:alpha_vs_E}(b){]}.
In other words, the optimal $\alpha$ for temperatures of 380 and
500 K (which is characteristic of the maximal field-induced change
of entropy at these temperatures) is accompanied by subtle local rearrangements
of the dipolar pattern in PMN. We also numerically checked (not shown
here) that $\alpha_{max}$ at even higher temperature, such as 700
K, is also associated with such aforementioned local features involving
dipoles lying near the {[}1$\overline{1}$1{]}, {[}11$\overline{1}${]}
or {[}$\overline{1}$11{]} pseudocubic directions. Note also that
correlation between enhancement of EC coefficients and occurrence
of local features was found in the lead-free BZT relaxor ferroelectric
too \cite{Jiang2017}, except that the precise local quantity associated
with $\alpha_{max}$ is different between BZT and PMN \textendash{}
likely because the field was applied along {[}001{]} rather than {[}111{]}
in our previous study about EC coefficient in BZT \cite{Jiang2017}.
As a matter of fact, the dipoles involved in the local features of
BZT inherent to the enhancement of $\alpha_{max}$ are those pointing
near the four <111> pseudocubic directions having a positive z component,
that are {[}111{]}, {[}$\overline{1}$11{]}, {[}1$\overline{1}$1{]}
and {[}$\overline{1}$$\overline{1}$1{]}.

\section{Summary}

In summary, we employed the effective Hamiltonian of Ref. {[}\onlinecite{Prosandeev2015}{]}
to shed some light on electrocaloric effects in PMN. It is particularly
striking that such Hamiltonian can qualitatively reproduce not only
the peculiar electric field-\textit{versus}-temperature phase diagram
but also the optimization of the EC coefficient near the critical
point in this rather complex system. The fact that the recently developed
Landau-like model, predicting that the EC coefficient is simply related
to the product of temperature and the field derivative of the square
of the polarization \cite{Jiang2017}, also describes well the EC
behavior of PMN as a function of electric field and temperature is
also promising for phenomenological modelization of complex inhomogeneous
systems. Moreover, we hope that the present discoveries that the giant
EC coefficient in the vicinity of the critical point corresponds to
the percolation threshold while (the still large) $\alpha_{max}$
for higher temperatures is related to other specific microscopic features
further lead to a better understanding of EC effects and relaxor ferroelectrics.
It will also be interesting in a near future to investigate the effect
of long-range and/or short-range chemical orders between Mg and Nb
ions on the electrocaloric response of PMN, since properties of such
system has been shown to be dependent on it \cite{Sergey2016}.
\begin{acknowledgments}
Z.J., S. Prokhorenko, and L.B. are grateful for the DARPA Grant No.
HR0011-15-2-0038 (MATRIX program) for support. Z.J. also acknowledges
support from the National Natural Science Foundation of China (NSFC),
Grants No. 11574246, No. 51390472, and No. U1537210, National Basic
Research Program of China, Grant No. 2015CB654903, and China Scholarship
Council. Y.N. is supported by ARO Grant No. W911NF-16-1-0227. S. Prosandeev
is supported by ONR Grants No. N00014-12-1-1034 and N00014-17-1-2818,
and Grants No. 3.1649.2017/4.6 from RMES (Russian Ministry of Education
and Science), and No. 18-52-0029 Bel\_a from RFBR (Russian Foundation
for Basic Research). We also acknowledge funding from the Luxembourg
National Research Fund through the intermobility (Grant No. 15/9890527
Greenox, J.Í. and L.B.) program. Some computations were also made
possible owing to MRI Grant No. 0722625 from NSF, ONR Grant No. N00014-15-1-2881
(DURIP), and a Challenge grant from the Department of Defense. 
\end{acknowledgments}

\end{document}